\documentclass[10pt,aps,prd,twocolumn,showpacs,amsmath,amssymb,longbibliography,nofootinbib,preprintnumbers]{revtex4-1}
\usepackage{graphicx}
\usepackage{subfigure}
\usepackage{amssymb}
\usepackage{slashed}
\usepackage{hyperref}
\usepackage{dsfont}
\hypersetup{colorlinks=true,citecolor=blue}

\newcommand{\CO}{{\cal O}}
\newcommand{\CN}{{\cal N}}

\newcommand\rar{\rightarrow}

\newcommand{\N}[1]{\ensuremath{\mathcal N=#1}}

\begin{document}
 \title{A Precision Test of AdS/CFT with Flavor}

 \author{Andreas Karch}
 \email{akarch@uw.edu}
 \author{Brandon Robinson}
 \email{robinb22@uw.edu}
 \author{Christoph F.~Uhlemann}
 \email{uhlemann@uw.edu}

 \affiliation{Department of Physics, University of Washington, Seattle, WA 98195-1560, USA}

\begin{abstract}
In this letter we put AdS/CFT dualities involving probe branes to a precision test.
On the holographic side we use a new class of supersymmetric D7-brane embeddings into AdS$_5\times$S$^5$,
which allow to describe \N{4} SYM coupled to massive \N{2} supersymmetric flavors on S$^4$.
With these embeddings we can compare holographic results to a field theory analysis of the free energy
using supersymmetric localization.
Localization allows us to get results at strong coupling, and hence to compare in detail to AdS/CFT.
We find analytically matching results: a phase transition at the same critical mass in both calculations and
matching free energies up to a scheme-dependent constant in both phases.
\end{abstract}
\maketitle

\section{Introduction}
Probe branes have found a wide range of applications in holographic studies, as
the simplifications provided by the probe approximation make them a very versatile tool.
They are used, e.g., to add quarks to holographic duals of QCD-like theories \cite{SS} and give one of the
simplest holographic realizations of compressible and conducting matter \cite{AOB}.
Strictly speaking, the addition of probe branes is an extra ingredient in holography.
It does not directly follow from the basic postulates, and one may in addition be worried about the probe limit being well defined.
Conducting a decisive test of these dualities is tough, however.
The virtue of the dualities, i.e.\ that involved questions on one side
are mapped to simple ones on the other and vice versa, becomes an obstacle when it comes to testing.
It is just difficult to calculate the same quantity in the same regime on both sides of the dualities.
Building on recent progress in the study of supersymmetric gauge theories on
curved, compact manifolds, and in particular supersymmetric localization \cite{Pestun},
we give a detailed test in this work.

In \cite{Pestun}, a massive deformation of $\CN=4$ Super Yang-Mills (SYM) theory, called $\CN=2^*$, was constructed on S$^4$.
Preserving a subset of the supersymmetries allowed for the use of supersymmetric localization.
This procedure reduces the partition function from an infinite-dimensional path integral to an ordinary integral over a modified Gaussian matrix model.
This dramatic simplification makes exact calculations possible and has led to a large volume of work studying its application
in the context of AdS/CFT \cite{HEDFNBSP, RZB, AB}.
In particular, the authors of \cite{HEDFNBSP} constructed, albeit numerically, the gravitational dual to $\CN=2^*$ on an S$^4$, and were able to perform a rigorous test of AdS/CFT by matching derivatives of free energies.

The methods of \cite{Pestun} can be applied to more general $\CN=2$ supersymmetric gauge theories on S$^4$.
Of particular interest are QCD-like theories with $N_c$ colors and $N_f$ matter multiplets in the fundamental representation of the gauge group.
In the limit of large $N_c$, fundamental matter offers a new small parameter, $\zeta \equiv N_f/N_c$.
The matter fields experience non-trivial dynamics in the background of the gauge field even in the $\zeta \rightarrow 0$ limit.
However, there are not enough matter degrees of freedom to alter the dynamics of the color fields.
That simplification is captured holographically by the probe limit \cite{KK}.
The fundamental flavor multiplets get incorporated via a brane that minimizes its action in a fixed background geometry.
Its backreaction can be neglected.
Building on our recent construction of supersymmetric probe brane embeddings dual to $\CN=4$ SYM coupled to massive fundamental
matter on curved spaces \cite{AKBRCU}, we are now in a position to perform a precise check of this theory using localization,
with both sides of the correspondence under complete analytic control.

Even for a gauge theory on a compact manifold,
there can be non-trivial phase structure owed to large $N_c$ \cite{DGEW}.
Including fundamental matter raises interesting puzzles on the field theory side,
where localization calculations with fundamental matter hint at a complicated and sometimes poorly understood phase structure \cite{RZ1}.
In the theory we are studying we have complete control over the localization calculation and can identify a single well-characterized phase transition as a function of mass.

In Sec.~\ref{sec:holography}, we start with the holographic side.
We discuss the brane embedding and evaluate the chiral and scalar condensates as well as $dF/dM$, where $M$ is the mass of the flavors and $F$ the free energy.
In Sec.~\ref{sec:localization}, we turn to the localization computation.
We discuss the quenched approximation of the matrix model and also calculate $dF/dM$,
to compare to the holographic result.
We end in Sec.~\ref{sec:discussion} with a discussion.

\section{Holographic Probe Brane Analysis}\label{sec:holography}
The essential ingredient to finding supersymmetric brane embeddings is to preserve $\kappa$-symmetry.
This is an extra fermionic gauge symmetry used to project out part of the fermionic modes, such as to obtain
matching numbers of bosonic and fermionic degrees of freedom \cite{EBPT,Cederwall:1996pv,Cederwall:1996ri,EBRKTOGP}.
We discuss the brane embeddings first and then the computation of one-point functions and free energy.

\subsection{Supersymmetric Embeddings}
To describe the dual theory on S$^4$, we start with a spherically-sliced global AdS$_5\times$S$^5$ background
in Euclidean signature.
In Fefferman-Graham gauge the line element reads
\begin{subequations}
\begin{align}
ds^2&= \frac{dz^2}{z^2}+\frac{(1-\frac{z^2}{4})^2}{z^2}d\Omega_4^2+d\Omega_5^2~,\\
d\Omega_5^2&=d\theta^2+\sin^2\theta d\Omega_3^2+\cos^2\theta d\psi^2~.
\end{align}
\end{subequations}
The D7-branes, described by the action given in (\ref{eqn:D7-action}), are embedded into this background, and the embedding is characterized
in static gauge by the slipping mode $\theta(z)$ alone.
The induced metric reads
\begin{align}
ds_{D7}^2&=\frac{1+z^2{\theta^\prime}^2}{z^2}dz^2+\frac{(1-\frac{z^2}{4})^2}{z^2} d\Omega_4^2+\sin^2\theta d\Omega_3^2~.
\end{align}
The asymptotic D7-brane geometry is AdS$_5\times$S$^3$, and the profile of $\theta(z)$ determines whether and where
the branes cap off via the internal cycle collapsing.
$\CN=2$ supersymmetric field theories on $S^4$ with massive fields require the addition of a dimension-2 scalar-bilinear compensating term in the Lagrangian,
in order to restore the supersymmetry that is otherwise broken by the curvature \cite{Pestun, GFNS}.
To source these compensating terms holographically, we need to turn on a worldvolume 1-form gauge field on the D7, $A=f(z)\omega$.
To reflect the properties of the field-theory mass term, $A$ has to transform in a specific way under the SO(4) isometries of the S$^3$
that the D7-branes wrap in the internal space.
In the language of \cite{MKDMRM}, this translates to $\omega$ transforming as $(0,\,1)$ under $SU(2)\times SU(2)$.

We now turn to the $\kappa$-symmetry analysis.
Our starting point is the fact that the brane embedding preserves those supersymmetries of the background geometry,
which are generated by Killing spinors that satisfy a projection condition, $\Gamma_\kappa\epsilon=\epsilon$.
The matrix $\Gamma_\kappa$ encodes the brane embedding.
To find supersymmetric embeddings, we turn the logic around.
Feeding in the explicit form of the AdS$_5\times$S$^5$ Killing spinors and demanding that there be non-trivial solutions
to the projection condition, we find a set of necessary conditions on the embedding and worldvolume flux.
This is technically involved.
We give the details in  \cite{AKBRCU} and content ourselves with an outline of the main points here.
The projection condition as given in \cite{EBPT}, in Euclidean signature and with our ansatz for the gauge field and embedding, takes the form
\begin{align}
\left(\mathds{1}+\frac{1}{8}\gamma^{ijkl}F_{ij}F_{kl}\right)\hat{\Gamma}\epsilon+\frac{1}{2}\gamma^{ij}F_{ij}\hat{\Gamma}C\epsilon^\star&=h\epsilon~.
\end{align}
Here $\gamma_i$ denotes the pullback of the spacetime gamma matrices to the D7 worldvolume,
$\hat{\Gamma}$ is a $\theta$-dependent linear combination of $\gamma_i$ structures,
$h$ is the D7 brane DBI Lagrangian, and $C$ plays the role of charge conjugation.

Using the explicit form of the AdS$_5\times$ S$^5$ Killing spinors given in \cite{AKBRCU} and following the
logic outlined above
fixes $\omega$ to be precisely what we argued for,
and gives us a non-linear relation between the gauge field and slipping mode.
In addition, we find a 2$^\mathrm{nd}$-order differential equation for the slipping mode alone.
That equation can be solved analytically, and we find
\begin{subequations}\label{eqn:brane-embedding}
\begin{align}
\cos\theta(z)&=2\cos\left(\frac{2\pi k+\cos^{-1}\tau(z)}{3}\right)~,\\
\tau(z)&=\frac{96z^3(c-m\log\frac{z}{2})+6mz(z^4-16)}{(z^2-4)^3}~,\\
f(z)&=-i\sin^3\!\theta\,\frac{z(z^2-4)\theta^\prime-(z^2+4)\cot\theta}{8z}~.\label{eqn:gauge-field-from-slipping-mode}
\end{align}
\end{subequations}
The values $k\in\{0,\,1,\,2\}$ correspond to different branches of $\cos^{-1}$, which we take to be valued in $[0,\pi]$.
We will use $k\,{=}\,2$ such that $\theta$ is real at the boundary.
The parameter $m$ is identified, up to a factor of the tension of a fundamental string,
with the mass of the flavor fields as $M=m\sqrt{\lambda}/2\pi$ \cite{MKDMRM}.
The normalization factor $\mu\equiv\sqrt{\lambda}/(2\pi)$ will be crucial in the field theory analysis:
for any $m$ which is not infinitesimally small, we will deal with heavy flavors in the field theory.
That is, our quarks have mass of order $\sqrt{\lambda}$ in units of the S$^4$ radius.
$c$ appears in the condensate $\langle\bar{\psi}\psi\rangle$ and controls its non-analytic behavior.
The relation (\ref{eqn:gauge-field-from-slipping-mode}) in particular links the near-boundary expansions of $f$ and $\theta$,
which should be expected
given that, on the field theory side, the coefficient of the compensating term is fixed by the superpotential mass \cite{Pestun}.

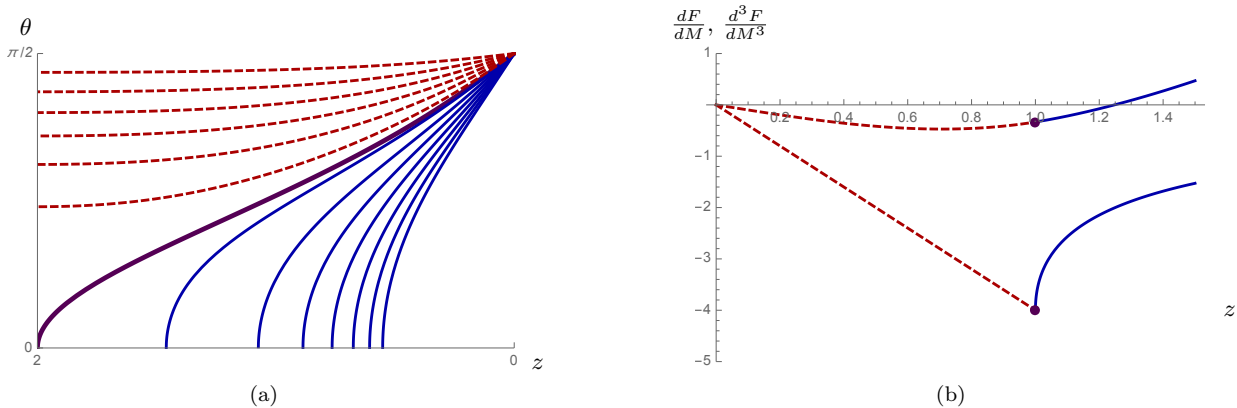
\begin{figure*}[htb]
\center
\subfigure[][]{ \label{fig:Theta}
  \includegraphics[width=0.38\linewidth]{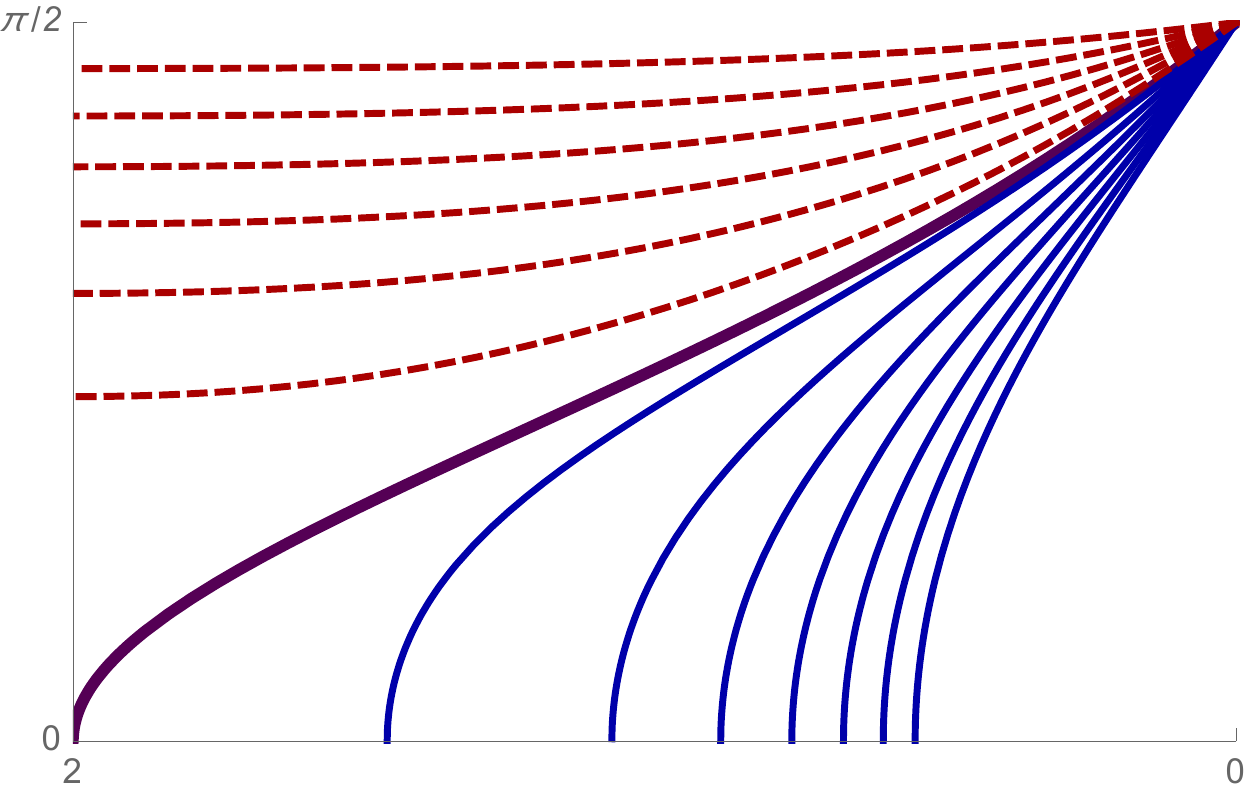}
}\hspace*{20mm}
\subfigure[][]{ \label{fig:F3}
  \includegraphics[width=0.38\linewidth]{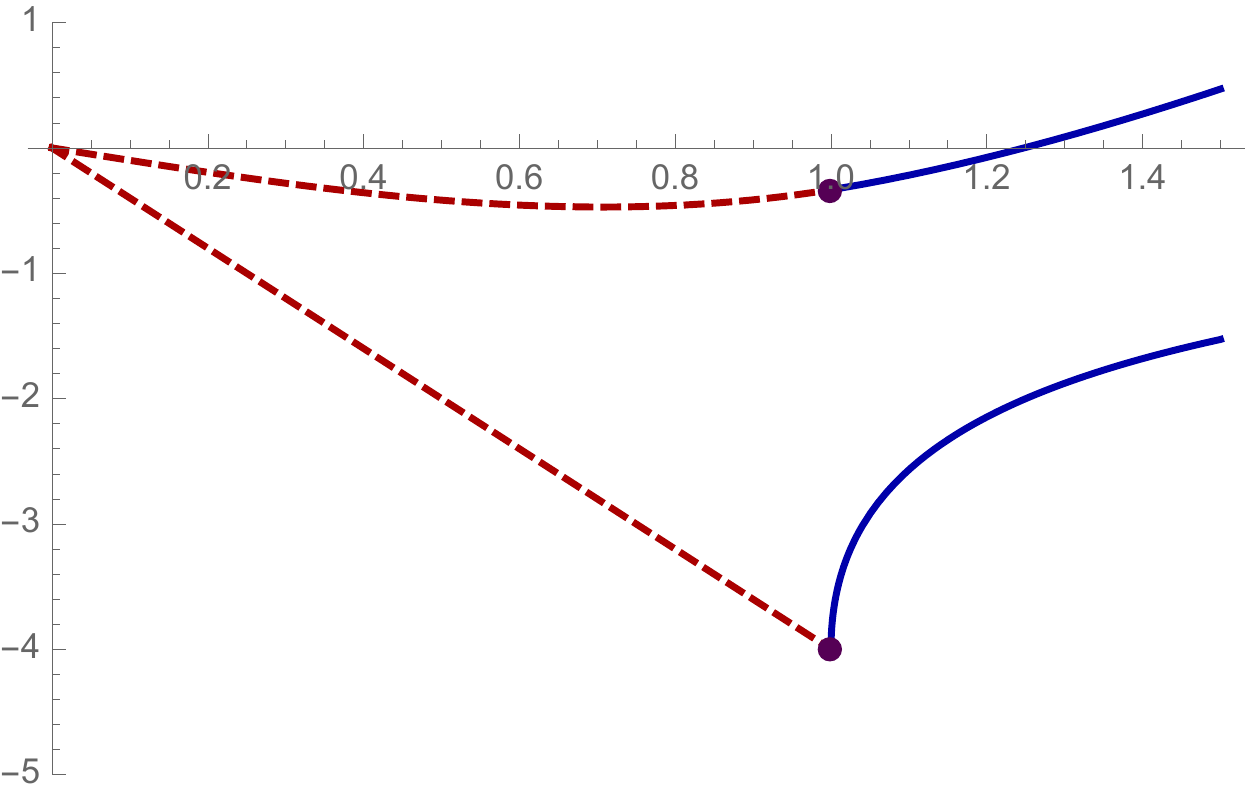}
}
\begin{picture}(0,0)
\setlength{\unitlength}{1cm}
\thicklines
\put(-16.0,4.5){$\theta$}
\put(-9.2,0.05){$z$}
\put(-7.35,4.55){$\frac{dF}{dM}$, $\frac{d^3F}{dM^3}$}
\put(-0.0,0.75){$z$}
\end{picture}
\caption{
On the left hand side the slipping mode is shown for $m\in\lbrace 0, \,0.025,\,\ldots,2 \rbrace$.
The dashed red curves ending on the vertical axis are embeddings where the S$^4$ collapses at $z\,{=}\,2$.
The solid blue curves ending on the horizontal axis are embeddings where the brane caps off
at a $z_\star\in(0,2)$ with the internal S$^3$ collapsing.
The critical embedding with $m=1$, where the S$^3$ and S$^4$ collapse concurrently, is shown as the thick purple curve ending at the origin.
The upper and lower curves on the right hand side show $dF/dM=V_4 \langle O_s\rangle$ 
and $d^3 F/dM^3$, respectively.
The linear term in $dF/dM$ is scheme dependent, and we subtracted off $2m\log\frac{\mu}{2}$ for the plot.
Since the matrix model and holographic calculations match analytically, we only plot the curves once.
The color/line coding reflects the kind of embedding from which the results are obtained holographically.
In the matrix model calculation, the red dashed parts are obtained from (\ref{eq: Fless})
and the blue solid parts from (\ref{eq: Fgreat}). They meet at the purple dots, which
correspond to the hyper moving on to the eigenvalue distribution.
\label{fig:D7-embeddings}}
\end{figure*}

The D7 brane embeddings come in two distinct classes: the branes can either smoothly cap off at a $z_*\in(0,2)$,
or they can extend all the way to the center of AdS at $z=2$.
The D7-brane geometry is a cone with an S$^3$ $\times$ S$^4$ base, where the S$^3$ lives in the internal space and the S$^4$ is the radial slice in AdS.
For the first type of embeddings, the $S^3$ shrinks at the tip of the cone, whereas for the second it is the $S^4$ that shrinks.
These two types of embeddings are connected by a critical embedding where the brane caps off at $z_*=2$.
In that case the spheres simultaneously collapse at the tip.
The condition that must be satisfied for a brane to cap off at a $z_\star\in(0,2)$ is $\theta(z_*)\in \lbrace 0,\,\pi\rbrace$,
which determines $c$ as
\begin{subequations}\label{eqn:pt1pt2}
\begin{align}\label{pt1}
c&=\frac{96mz_*^3\log\frac{z_*}{2}-6mz_*(z_*^4-16)\pm(z_*^2-4)^3}{96z_*^3}~.
\end{align}
The gauge field configuration at $z=z_*$ is singular unless $f(z_*)=0$, which fixes the cap off point in terms of the mass as
\begin{align}\label{pt2}
z_*=2(m-\sqrt{m^2-1})~.
\end{align}
\end{subequations}
Note that these capped embeddings only exist for $m>1$.
For $m<1$, including the case of massless flavors, we instead find embeddings that fill all of AdS.
A smooth embedding in that case requires $\theta^\prime(z=2)=0$, which translates to $c=0$.
These two topologically distinct families merge, as we will see, in a continuous phase transition at $m=1$.
For a plot of the corresponding slipping modes see Fig.~\ref{fig:Theta}.

\subsection{One-Point Functions and Free Energy}
With the embedding in hand, the computation of CFT one-point functions follows
the standard AdS/CFT prescription, and we give the details in App.~\ref{app:holo-computation}.
We can vary the asymptotic values of $\theta$ and $f$ independently, and thus calculate the
chiral condensate $\CO_\theta$ and the scalar condensate $\CO_f$ individually.
However, if we insist on staying within the family of supersymmetric embeddings,
the variations are related, and we only get a particular linear combination $\CO_s\equiv\CO_\theta+i\CO_f$.
This is the only expectation value we will have access to in the localization calculation.
Varying the D7-brane action with respect to the field theory mass $M=m\mu$, we find
\begin{align}\label{eqn:susy-condensate}
\frac{\mu}{T_0} \langle\CO_s\rangle &=3c+\frac{{2m}^3}{3}(1+6\alpha_1)-\frac{m}{2}(7+4\beta)~,
\end{align}
where $T_0=T_7V_{S^3}$ and $\alpha_1,\,\beta$ are scheme-dependent coefficients of finite counterterms.
That is, they are ambiguities in the renormalization procedure.
Demanding the renormalization scheme to preserve supersymmetry on flat space/Poincar\'e AdS fixes
$\alpha_1\,{=}\,-\frac{5}{12}$ \cite{AKAoBKS}.
To translate $T_0$ in (\ref{eqn:susy-condensate}) to field theory quantities, we use (see e.g.\ the table in \cite{HCAK})
\begin{align}\label{eqn:T0-lambda}
 T_0 V_4/N_c^2 &=\lambda\zeta/6\pi^2 = 2\mu^2\zeta/3~,
\end{align}
where $V_4$ denotes the volume of the unit S$^4$.
This in particular results in a free energy proportional to $\lambda$ at strong coupling,
which has long been recognized as a puzzling feature of the probe brane analysis,
and the localization calculation will have to reproduce that.
Note that $V_4\langle\CO_s\rangle=dF/dM$,
so (\ref{eqn:susy-condensate}) with (\ref{eqn:pt1pt2}), (\ref{eqn:T0-lambda}) can be readily compared to the field theory side.
For a plot see Fig.~\ref{fig:F3}.

On the matrix model side, analyses of massive large-$N_c$ $\CN=2$ gauge theories on S$^4$ have seen peculiar, infinite in multitude,
phase transitions as one takes the decompactification limit at strong coupling, when more and more resonances are excited on the eigenvalue
distribution \cite{RZ1,RZ2}.
In our setup on the probe brane side, we see exactly one, topology changing, transition between the phases where we have spacetime filling branes
for $m\,{<}\,1$ and branes that cap off smoothly for $m\,{>}\,1$.
The (quantum) critical point occurs exactly at $m\,{=}\,1$, where the wrapped S$^3\subset$ S$^5$ collapses concurrently with the S$^4$ at the origin.
To determine the critical exponent we expand (\ref{eqn:susy-condensate}) around the critical embedding.
For $m=1+\epsilon$ with $\epsilon\ll 1$, we find
\begin{align}\label{eq:GravExp}
\begin{split}
\frac{V_{4}}{\zeta\mu N_c^2} \langle\CO_s\rangle\simeq& -\frac{27+12\beta}{9}-\frac{13+4\beta}{3}\epsilon
-2\epsilon^2
\\&
+\frac{16\sqrt{2}}{15}\epsilon^{5/2}
+\CO(\epsilon^3)~.
\end{split}
\end{align}
The striking feature of this expansion is that we have full analytical control over extracting the critical exponents for this manifestly quantum phase transition.
It should be mentioned that these are distinct from the study of non-SUSY flavors in \cite{AKAoBLY}. The difference can be traced back
to the imaginary gauge field which gives non-trivial cancellations in the action that modify the general scaling analysis of \cite{AKAoBLY}.

\section{Localization with Quenched Flavors}\label{sec:localization}

Before we derive $dF/dM$ on the field theory side, we review where the components in the matrix model originate.
The localization calculation \cite{Pestun} begins by identifying a Grassmann scalar symmetry, $Q$, that is nilpotent up to gauge transformations.
This procedure requires closure of the supersymmetry algebra off shell, which needs an appropriate set of auxiliary fields.
After adding a $Q$-exact term $\delta_V=tQV$ to the Lagrangian, to which the partition function is insensitive, one can take the limit
$t\rar\infty$ and study the saddles of the path integral where $\delta_V$ vanishes.
The partition function reduces to an integral over the locus in field space where $\delta_V=0$.

In the study of $\CN=2$ gauge theories on S$^4$, the locus ends up being where all of the fields vanish, except for one constant adjoint-valued scalar.
Computing the 1-loop fluctuations about the locus exactly determines the partition function, up to instanton corrections.
The latter are exponentially suppressed in the large $N_c$ limit \cite{NNAO}, and so we ignore them here.
For $\CN=4$ SYM, the 1-loop determinants evaluate to unity, and one ends up solving a simple unitary Gaussian matrix model
\begin{align}
Z&=\int\! da^{N_c-1}\prod_{i<j} a_{[ij]}^2\, e^{S^{}_0}\,,&
S_0&={-\frac{8\pi^2}{\lambda}N_c\sum_i a_i^2}\,,
\end{align}
where $a_{[ij]}=a_i-a_j$ labels the roots of $\mathfrak{su}(N_c)$ and $a_i$ labels the weights.
The Vandermonde determinant factor $\prod_{i<j}a_{[ij]}^2$ comes from gauge fixing into the Cartan subalgebra.
It provides a repulsive logarithmic interaction term for the eigenvalues.

For any $\CN=2$ gauge theory on S$^4$ with massive hypermultiplets, the 1-loop fluctuations can be encoded in the following mnemonic:  
for each vector multiplet, adjoint hyper, and fundamental hyper we acquire a 1-loop factor
\begin{eqnarray}
&&\text{Vector}:\quad \prod_{i<j}H^2(a_{[ij]})\,,\\
&&\text{Adjoint}:\quad \prod_{i<j}\frac{1}{H(a_{[ij]}+M_A)H(a_{[ij]}-M_A)}\,,\\
&&\text{Fundamental}:\quad \prod_{i,f} \frac{1}{\sqrt{H(a_i+M_f)H(a_i-M_f)}}\,,~~
\end{eqnarray}
where $H(x) = G(1+ix)G(1-ix)$ and $G(x)$ is the Barnes G-function, $M_A$ is the adjoint hyper mass,
and $M_f$ are independent flavor masses indexed by $f$.\footnote{This counting scheme for fundamental hypers differs from \cite{RZ1,RZ2,XCJGKZ},
where the authors count `fundamental' and `anti-fundamental' hypers with $\prod_i H^{-1}(a_i\pm M)$, respectively,
suggestive of $\CN=1$ chiral and anti-chiral multiplets.}
For notational ease, we will denote $H(x\pm M)\equiv H_\pm(x)$ and $x_\pm = x\pm M$.
The result for $\CN=4$ SYM coupled to massive \N{2} flavors is a matrix model given by
\begin{align}
Z&= \int d^{N_c-1}a\, \frac{\prod_{i<j}a_{[ij]}^2}{\prod_i \sqrt{H^{N_f}_+(a_i)H^{N_f}_-(a_i)}}\,e^{S^{}_0}~.
\end{align}
Rearranging such that all factors appear in the exponent, we get an integrand $e^S$ with
\begin{align}
  S&=S_0-N_c\sum_i\frac{\zeta}{2}\log (H_+H_-)+\sum_{i<j}\log a_{[ij]}^2~.
\end{align}
The quenched approximation then becomes a matter of evaluating the partition function with $1\,{\ll}\, N_f\,{\ll}\, N_c$.
Note that the sums $\sum_i$ and $\sum_{i<j}$ are $\CO(N_c)$ and $\CO(N_c^2)$, respectively.
The calculation of the free energy  can be organized according to an expansion in $\zeta$ by using 
$F\approx -S|_\mathrm{saddles}$ and
\begin{align}
\frac{S}{N_c^2} &= \tilde{S}_{0}|_{\rho_0} + \zeta(S_1|_{\rho_0}+\delta \tilde{S}_0|_{\rho_0})+\CO(\zeta^2)~,
\end{align}
where $N_c^2\tilde{S}_0=S_0+\sum_{i<j}\log a_{[ij]}^2$, and $S_1$ is the contribution of the flavors.
Here we have denoted the solution to the Gaussian matrix model corresponding to pure $\CN=4$ SYM in the continuum limit as $\rho_0$,
which is the Wigner semicircle distribution
\begin{align}
\rho_0 (x)&= \frac{2}{\pi\mu^2}\sqrt{\mu^2-x^2}~,
\end{align}
where $\mu = \sqrt{\lambda}/2\pi$ is the maximum eigenvalue.
Note that, since $\rho_0$ extremizes $\tilde S_0$, $\delta \tilde{S}_0|_{\rho_0}=0$.
Thus, our analysis only requires the knowledge of $S_1|_{\rho_0}$.

Since we want to compare to AdS/CFT, in addition to $\zeta\,{\ll}\, 1$ we need to be working in the strong coupling, $\lambda\,{\gg}\, 1$, limit.
The masses of our flavors are of order $\sqrt{\lambda}$ and the typical eigenvalue contributing to our integral is also of
order $\mu \sim \sqrt{\lambda}$.
So the arguments of $H$ are large, validating the use of the asymptotic expansion of the log
derivatives
\begin{align}
H^\prime(x_\pm)/H(x_\pm)&= -x_\pm\log x_\pm^2+ 2x_\pm+\CO(x_\pm^{-1})~.
\end{align}
When $|M|\,{<}\,\mu$ and the hypers lie on the eigenvalue distribution,
using the semicircle distribution and large-argument expansion is only justified outside of a region of width $1/\sqrt{\lambda}$ around $x=M$,
where the hypers are parametrically light.
But the contribution of that region is negligible at large $\lambda$.
Consequently, the flavor contribution is
\begin{align}\label{eqn:Fprime-matrixmodel}
F^\prime&=\frac{\zeta N_c^2}{2}\!\int\limits_{-\mu}^\mu \!dx\,\rho_0(x)\left[4M-x_+\log x_+^2 + x_-\log x_-^2\!\right],
\end{align}
where $F^\prime=dF/dM$.
Integrating explicitly in the regime where $\mu<M$, we find
\begin{align}
\begin{split}\label{eq:tomatch}
F^\prime=&\,\frac{\zeta N_c^2}{3\mu^2}\Big[-2M^3+2\sqrt{M^2-\mu^2}(M^2+2\mu^2)\\
&~~+3M\mu^2\Big(1-2\log\frac{M+\sqrt{M^2-\mu^2}}{2}\Big)\Big].\,\,\,
\end{split}
\end{align}
In the matrix model, phase transitions can occur when some of the hypers become light,
as demonstrated e.g.\ in $\CN=2^*$ in \cite{RZ2}.
That is, as we take the hyper mass on to the eigenvalue distribution, $M\leq \mu$,
there can be resonances driving the hypers effectively massless.
Zooming in on the potential phase transition point, $M=m\mu$, with $m=1+\epsilon$ and expanding for $\epsilon \,{\ll}\, 1$, we find
\begin{align}\label{eq:MMExp}
\begin{split}
\frac{dF/dm}{N_c^2\mu^2\zeta}
=&\,
\frac{1}{3}-\log\frac{\mu^2}{4}-\epsilon\Big(1+\log\frac{\mu^2}{4}\Big)\\
&-2\epsilon^2+\frac{16\sqrt{2}}{15}\epsilon^{5/2}
+\CO(\epsilon^3)~.
\end{split}
\end{align}
This expansion reproduces the non-analytic behavior and matches exactly the coefficients of all the $\epsilon^{(2n+1)/2}$ terms that were
seen on the gravitational side in (\ref{eq:GravExp}), which is strong evidence that the topology changing phase transition is captured by
the transition associated with bringing hypers on to the distribution.

If we set the remaining scheme dependent counterterm to $\beta\,{=}\,-\frac{5}{2}+\frac{3}{2}\log\frac{\mu}{2}$, we can achieve an exact match
of the holographic result (\ref{eqn:susy-condensate}) with (\ref{eqn:pt1pt2}), (\ref{eqn:T0-lambda}) to (\ref{eq:tomatch}) for $\mu\,{<}\, M$:
\begin{eqnarray}\label{eq: Fgreat}
\frac{V_4}{\zeta \mu N_c^2}&& \langle\CO_s\rangle=-\frac{2m^3}{3}+\frac{2}{3}\sqrt{m^2-1}(m^2+2)\\\nonumber
 &&+m\left[1+2\log\frac{2(m-\sqrt{m^2-1})}{\mu}\right] = \frac{F^\prime}{\zeta \mu N_c^{2}} \,.
\end{eqnarray}
Since we have a good holographic description also of $m\,{<}\,1$ embeddings,
we should be able to perform the same calculation of $dF/dM$ also in the matrix model
for $M\,{<}\,\mu$. That is, even when the hyper mass is on the eigenvalue distribution. We indeed find again perfect agreement in this regime:
\begin{align}\label{eq: Fless}
V_4\langle\CO_s\rangle&= m\mu\zeta N_c^2\left(1-\frac{2}{3}m^2-2\log\frac{\mu}{2}\right)= F^\prime~.
\end{align}
The details of this calculation are given in App.~\ref{app:matrix-model}.

\section{Discussion}\label{sec:discussion}
We have studied the phase structure of \N{4} SYM coupled to massive \N{2} flavor hypermultiplets on S$^4$,
using holography and direct QFT computations independently.
The crucial ingredients to allow for localization of the path integral on the field theory side are the
preservation of some supersymmetry and the formulation on a compact space.
Holographically, this translates to the supersymmetry of the D7-brane embeddings we derived in \cite{AKBRCU}.
We found one continuous phase transition at the same value of the flavor mass in both calculations,
and analytically matching $dF/dM$ in both phases.
The remaining constant, which enters when this relation is integrated to get the free energies, is scheme dependent.
So matching the free energies themselves then merely amounts to choosing compatible renormalization schemes.

Our results give strong support to the validity of the probe brane constructions used so frequently in AdS/CFT.
The theories we studied are non-conformal, and the quantities we compared are not special,
in the sense that they are not extrapolated from weak to strong coupling using non-renormalization theorems.
Moreover, the theory described by the D3/D7 setup has a non-trivial UV fixed point only in the quenched approximation,
which frequently means that extra care is needed when establishing the validity of holographic results.
The fact that we found such nicely matching results therefore truly provides a non-trivial test of the dualities.

Possible directions for future research include tests for other probe brane systems like D3/D5 \cite{KR},
using localization on S$^4$ with defect hypers \cite{ODDFHO,JLP}, or the computation of superconformal indices. The crucial ingredient on the holographic side in either case will be to
find the corresponding supersymmetric embeddings. It would also be desirable to further investigate why other massive $\CN=2$ theories see curiously rich phase structures.

\section*{Acknowledgments}
We thank Tyson Price for contributions in the early phase of this work.
The work of AK and BR was supported, in part, by the US Department of Energy under grant number DE-SC0011637.
CFU is supported by {\it Deutsche Forschungsgemeinschaft} through a research fellowship.

\appendix

\section{Holographic renormalization and one-point functions}\label{app:holo-computation}

Making contact with CFT data requires computing a finite on-shell action employing holographic renormalization \cite{deHaro:2000xn, MBDFKS, AKAoBKS}.
The on-shell action diverges near the conformal boundary.
It can be regulated by introducing a radial cut-off $z\geq \delta$, and renormalized by then introducing a set of covariant counterterms at $z=\delta$.
The chiral (scalar) condensates, $\CO_\theta$ ($\CO_f$), are then computed by varying the on-shell action w.r.t.\ the boundary values of $\theta$ ($f$).
We lay out the technical details in this appendix.

We start with the D7-branes described by the DBI action with Wess-Zumino term
\begin{align}\label{eqn:D7-action}
\begin{split}
S_\mathrm{D7}=&-T_7\int d^8x\sqrt{\det\big[g+2\pi\alpha^\prime F\big]}\\
&+2(2\pi\alpha^\prime)^2\int_{\Sigma_8}C_4\wedge F\wedge F~,
\end{split}
\end{align}
where $g$ is the induced metric on the branes, $F$ is the worldvolume gauge field, $C_4$ is the RR 4-form potential,
and we absorb $2\pi\alpha^\prime$ into a redefinition of the gauge field.
The covariant counterterms needed for the slipping mode at $z=\delta$ are \cite{AKAoBKS}
\begin{align}
L_1&=-\frac{1}{4}\left[1-\frac{R}{12}+\frac{\log\delta}{8}\left(R_{ij}R^{ij}-\frac{R^2}{3}\right)\right]~,\\
L_2&=\frac{1}{2}\left(-\tilde{\theta}\Box_W\tilde{\theta}\log\delta+\tilde{\theta}^2\right)~,\\
L_\mathrm{fin}&=\alpha_1\tilde{\theta}^4+\alpha_2\tilde{\theta}\Box_W\tilde{\theta}+\frac{\alpha_3}{32}\left(R_{ij}R^{ij}-\frac{R^2}{3}\right)\,,
\end{align}
where $\Box_W=\Box +\frac{R}{6}$ is the Weyl covariant Laplacian and $\tilde{\theta}=\theta-\pi/2$.
In addition to these, we need the following counterterms associated with the gauge field
\begin{equation}
L_3=\frac{f^2}{2\log\delta}\left(1+\frac{2\alpha_4}{\log\delta}\right).
\end{equation}
The renormalized action, $S_\mathrm{D7,ren}=S_\mathrm{D7}-S_\mathrm{ct}$, obtained by supplementing $S_\mathrm{D7}$ in (\ref{eqn:D7-action})
with these counterterms, is now finite as $\delta\rar 0$.
The chiral condensate is given by
\begin{equation}
\mu \langle\CO_{\theta}\rangle= -\frac{1}{\sqrt{g_{S^4}}}\frac{\delta S_\mathrm{D7,ren}}{\delta\theta^{(0)}}~.
\end{equation}
The factor of $\mu$ on the left hand side accounts for the fact that the coefficient of the $\mathcal O(z)$ term in the slipping mode,
$\theta^{(0)}=m$, is related to the actual source of the fermion bilinear, $M$, by a factor of $\mu$.
The computation of the scalar condensate proceeds analogously.

We now spell out the variations of (\ref{eqn:D7-action}).
After the variation is carried out, we can use identities following from the $\kappa$-symmetry analysis in \cite{AKBRCU}
to simplify the result, which yields
\begin{align}
\frac{\delta_\theta S_{D7}}{T_0 V_4} &= \int_{{z}_*}^\delta dz \partial_{z}\left(z^2\xi^\prime\sin^4\theta\,\theta^\prime\delta\theta\right)~,\\
\frac{\delta_f S_{D7}}{T_0 V_4}& = \int_{{z}_*}^\delta dz\,\partial_{z}\left[(z^2\xi^\prime\sin^2\theta f^\prime+8\xi f)\delta f\right]~,
\end{align}
where $\xi^\prime = z^{-5}(1-z^2/4)^4$. The variation of the counterterms is straightforward and reads
\begin{eqnarray}
\nonumber
\delta_\theta L_\mathrm{ct}&=& T_0\,\delta\theta\left(\theta +\frac{\theta}{6}R\log\delta+4\alpha_1\theta^3-\frac{\theta}{3}\alpha_2R\right), \qquad\\
\delta_f L_{\mathrm{ct}}& =& T_0\,\delta f\left(\frac{f}{\log\delta}+\frac{2\alpha_4 f}{(\log \delta)^2}\right).
\end{eqnarray}
With these results in hand, we can compute the one-point functions.
The last ingredient is the asymptotic expansion of the solutions (\ref{eqn:brane-embedding}), which reads
\begin{eqnarray}
\tilde{\theta}&\simeq& m z -(c-\frac{{m}}{2}({m}^2+\frac{3}{2})){z}^3+m {z}^3\log \frac{z}{2}+\ldots\quad\ \ \\
f&\simeq&im{z}^2\log\frac{z}{2}-\frac{i}{3}\left(3c-m({m}^2+3)\right){z}^2+\ldots.\quad
\end{eqnarray}
We then find the condensates
\begin{eqnarray}
\nonumber
\langle\CO_\theta\rangle &=& \frac{T_0}{\mu}\left[2c-\frac{m}{2}(5+8\alpha_2-4\log 2)  +{m}^3(1+4\alpha_1)\right], \\
\langle\CO_f\rangle &=&\frac{i  T_0}{\mu}\left[\frac{{m}^3}{3}-c+m(1+2\alpha_4-\log 2)\right].
\end{eqnarray}
In $\langle\CO_s\rangle=\langle\CO_\theta\rangle+i\langle\CO_f\rangle$, only a linear combination of $\alpha_2$ and $\alpha_4$ appears,
and so we used $\beta \equiv 2\alpha_2 +  \alpha_4-\frac{3}{2}\log 2$ for the result quoted in the main text.

\section{Evaluation of the matrix model for small mass}\label{app:matrix-model}

We now give more details on the large argument expansion of the integrand of $dF/dM$ for $M\,{<}\,\mu$.
We start from the integral on the right hand side of (\ref{eqn:Fprime-matrixmodel})
\begin{align}
\frac{2F^\prime}{\zeta N_c^2}&=\int\limits_{-\mu}^\mu dx\,\rho_0(x)\left[4M-x_+\log x_+^2 + x_-\log x_-^2\right].
\end{align}
The last term in the integrand is just a constant multiplying the Wigner distribution, and fixed by its normalization.
The remaining part of the integrand can be split up into the $x_\pm\log x_\pm^2$ parts.
Setting then $M=m\mu$ and $x=(a\mp m)\mu$, we find
\begin{align}
\frac{2F^\prime}{\zeta N_c^2}&=4m\mu-I(m)+I(-m)~,
\end{align}
where we introduced the short hand
\begin{align}
 I(m)&=\frac{2\mu}{\pi}\int_{m-1}^{m+1}da \,a\log(a^2\mu^2)\sqrt{1-(a-m)^2}~.
\end{align}
These integrals can then be done analytically and we find
\begin{align}
\frac{1}{\zeta N_c^2}\frac{dF}{dM}&= \frac{m\mu}{3}(3-2m^2-6\log\frac{\mu}{2})~.
\end{align}

\bibliography{FlavorLoc}
\end{document}